\documentclass[aps,prl,twocolumn,preprintnumbers,amsmath,amssymb,floatfix]{revtex4}
\usepackage{graphicx}
\usepackage{dcolumn}
\usepackage{bm}

\newcommand{\dd}{\mathrm{d}}
\newcommand{\e}{\mathrm{e}}
\newcommand{\w}{\wedge}
\newcommand{\bbm}{\left(\begin{matrix}}
\newcommand{\ebm}{\end{matrix}\right)}
\newcommand{\beq}{\begin{eqnarray}}
\newcommand{\eeq}{\end{eqnarray}}
\makeatother

\newcommand{\sfrac}[2]{{\textstyle\frac{#1}{#2}}}

\newcommand{\be}{\begin{equation}}
\newcommand{\ee}{\end{equation}}
\newcommand{\beqa}{\begin{eqnarray}}
\newcommand{\eeqa}{\end{eqnarray}} 
\def\nn{\nonumber} \def \bea{\begin{eqnarray}} \def\eea{\end{eqnarray}}

\newcommand{\barr}{\begin{array}}
\newcommand{\earr}{\end{array}}

  \def\G{\Gamma}

\def\mc{\mathcal}

\def\one{\mbox{1 \kern-.59em {\rm l}}}

\def\bit{\begin{itemize}} 
\def\eit{\end{itemize}} 

\def\({\left(} \def\){\right)}

\begin{document}

\title{Generalized Yang-Mills theory and the D-brane effective action\footnote{ ITP-UH-09/14}}

\author{Athanasios Chatzistavrakidis$^1$}
\email{thanasis@itp.uni-hannover.de}
\author{Fridrik Freyr Gautason$^{1,2}$}
\email{fridrik.gautason@itp.uni-hannover.de}

\affiliation{${}^1$Institute for Theoretical Physics, Leibniz University Hannover \&\\ ${}^2$Center for Quantum Engineering and Spacetime Research, Appelstrasse 2, 30169, Hannover, Germany.
}

\date{\today}

\begin{abstract}
Non-abelian gauge theories in the context of generalized complex geometry are discussed.
  The generalized connection naturally contains
 standard gauge and scalar fields, unified in a
 purely geometric way. 
 We define the corresponding Yang-Mills theory on particular subbundles of a Courant 
 algebroid, known as Dirac structures, where the generalized curvature is a tensor.
 Different Dirac structures correspond to different known theories, such as 
the bosonic sector of maximally supersymmetric Yang-Mills 
in ten and four dimensions and reduced matrix models.
 Furthermore, we revisit 
 the non-abelian world volume effective action of D-branes in this formalism, where
 the gauge field on the 
brane and the transverse scalars are unified, while the action does not contain pullbacks of fields and its 
consistency with T-duality is verified at face value.
\end{abstract}


\maketitle

The D-brane effective action has the remarkable property that it keeps the same form under T-duality \cite{Myers:1999ps}. 
This is true for both 
the Dirac-Born-Infeld (DBI) sector, as well as the Wess-Zumino (WZ) sector that describes the couplings to the Ramond-Ramond (RR)
gauge potentials of the type II superstrings. The T-duality is realized 
on the world volume degrees of freedom,
 the gauge field $A_a$ and the transverse scalars $\Phi^i$, by exchanging their components in the dualized direction, i.e. 
\be \label{basic}
A_p\quad\overset{T_p}\longleftrightarrow \quad\Phi^p~.
\ee
On the other hand, the effective action depends on both the gauge and scalar fields, either explicitly or through pullbacks 
and interior products. Here we show that the two dual sets of degrees of freedom can be treated in a unified way in the context 
of generalized geometry, as components of a single generalized connection. This was already partially discussed in Ref. 
\cite{Satoshi}, for the abelian case of a single D-brane and for the DBI sector. Here we argue that this result extends to the non-abelian case of a 
stack of D-branes, and we examine the WZ sector. 

Our approach will be to start from the basic structures that appear in generalized complex geometry \cite{gg} in order to construct a non-abelian generalized gauge 
theory, where 1-forms and 1-vectors appear as components of a single generalized vector. We show that the corresponding 
generalized Yang-Mills (YM) action correctly reproduces theories with gauge and scalar fields, such as the bosonic sector of ${\cal N}=4$ supersymmetric YM theory and the D-instanton and D0 reduced matrix models, in a way 
that differs from standard dimensional reduction. 
Then we briefly review the proposal that a D-brane with its fluctuations is a twisted Dirac structure \cite{Satoshi,watamura2} (see also 
the discussion on generalized complex submanifolds
 in Ref. \cite{gg}, Refs. 
\cite{Grange:2005nm,Grange:2006es,Koerber:2005qi,Martucci:2005ht} for further work on this). This point of view will allow us to 
proove that the full non-abelian effective action for
  D-branes can be written solely in terms of a generalized curvature $\mc F$ 
and spacetime background fields.
 The resulting action is written without any pullbacks or interior products and its consistency 
with T-duality is verified at face value.

\section*{Generalized non-abelian gauge theory}
In this letter we will need some basic tools of
 generalized geometry \cite{Hitchin:2004ut,gg}. 
Let us consider a manifold $\text{M}$, equipped with local coordinates $x^P=(x^a,x^i),$ with $a=0,\dots,p$ and
 $i=p+1,\dots,m=\dim \text{M}-1$, and its generalized tangent bundle $\mc{T}\text{M}$, 
which is locally isomorphic to the sum of the tangent and cotangent bundle $\text{TM}\oplus\text{T}^{\star}\text{M}$. The space of sections 
$\G(\mc{T}\text{M})$ of the generalized bundle consists of generalized vectors $\mathfrak{X}=X+\eta$, where $X\in \G(\text{TM})$ is a standard 
1-vector and $\eta\in\G(\text{T}^{\star}\text{M})$ is a 1-form. The generalized tangent bundle is equipped with a bracket, called the Courant 
bracket and defined as 
\be 
\label{cour}[X+\eta,Y+\xi]_C=[X,Y]_{L}+{\cal L}_{X}\xi-{\cal L}_{Y}\eta-\frac{1}{2}\dd (\iota_X\xi-\iota_Y\eta)~,\nn
\ee
where the first term on the right hand side is the Lie bracket between two vectors, the middle terms are Lie derivatives of 
1-forms along vectors and the last term is the exterior derivative of interior products between vectors and 1-forms.
This structure can be promoted to a Courant algebroid \cite{wein}, provided that we introduce an anchor map  $\rho:\cal {T}\text{M}\to \text{TM}$, and a bilinear operation
\beq\label{bilinear}\langle X+\eta,Y+\xi\rangle=\frac{1}{2}(\iota_X\xi+\iota_Y\eta)~,\eeq
as well as certain axioms that provide compatibility conditions among them.
The Courant bracket is skew-symmetric but it does not satisfy the Jacobi identity at the level of the Courant algebroid. 
However, one can find subbundles with the property that the restricted Courant bracket is closed and satisfies the Jacobi identity, and 
where the bilinear operation gives zero for any two elements of the subbundle. Such subbundles are known as Dirac structures \cite{dirac}. 
Dirac structures are Lie algebroids and moreover the Courant algebroid can be identified with a Lie bialgebroid of the form 
$L\oplus L^{\star}$, where $L$ and $L^{\star}$ are two dual Dirac structures \cite{wein}.

According to Ref. \cite{gg}, a generalized connection (also called a Lie algebroid connection) is an operator $\mc D$ that maps the 
sections $\G(E)$ of a vector bundle $E$ to the sections $\G(L^{\star}\otimes E)$, where $L$ is a Lie algebroid, which we 
will consider here to be a Dirac structure. 
It satisfies a Leibniz rule,
\be
\mc D(fs)=(\dd_L f)\otimes s+f\mc D s~,
\ee 
where $f$ is an arbitrary smooth function on $M$, $s$ is a section of the vector bundle $E$, and $\dd_Lf=\partial_af\dd x^a$.
The generalized connection of interest for our purposes is
\be\label{gc}
\mc D=\dd_L+\mc A~,
\ee
where $\mc A=A+\Phi$ is a generalized vector, which consists of a 1-form $A=A_a\dd x^a$ (a standard gauge field) and a vector 
$\Phi=\Phi^i\partial_i$ (a multiplet of scalars fields). Moreover we consider the following dual pair of Dirac structures;  one
is identified with the subbundle 
 $L=\text{span}\{\partial_a,\dd x^i\}\subset\mc{T}\text{M}$, and the dual one is  $L^{\star}=\text{span}\{\partial_i,\dd x^a\}\subset \mc{T}\text{M}$. 
Obviously, $L\oplus L^{\star}\cong \mc{T}\text{M}$. Additionally, we  assume that the gauge field takes values 
in the Lie algebra of a non-abelian gauge group, thus implementing the non-abelian case in the present formulation. However, we suppress the corresponding gauge indices.
 
 The curvature $\mc F$ of a generalized connection $\mc D$ is defined as 
 \be \label{gcurvgen}
 \mc F(\mathfrak X,\mathfrak Y)=\mc D_{\mathfrak X}\mc D_{\mathfrak Y}-\mc D_{\mathfrak Y}\mc D_{\mathfrak X}-\mc D_{[\mathfrak X,\mathfrak Y]}~,
 \ee 
 where $\mathfrak X,\mathfrak Y\in \G(L)$. At the level of the Dirac structure, this curvature operator is tensorial in 
 nature as an element of $\G(\w^2L^{\star})$, which is not generally true for the full Courant algebroid \cite{Gualtieri:2007bq}. 
For this reason  it  makes sense to write a physical theory on a Dirac structure. Using the definition (\ref{gcurvgen}), it is simple 
 to compute the curvature of the particular generalized connection (\ref{gc}). We find:
 \bea
 \mc F&=&\sfrac 12 F_{ab}\dd x^a\w\dd x^b+(\partial_a\Phi^i+[A_a,\Phi^i])\dd x^a\w\partial_i+\nn\\&&+\sfrac 12[\Phi^i,\Phi^j]\partial_i\w\partial_j~,\label{gcurv}
 \eea 
 where $F=\dd A+A\w A$ is the non-abelian curvature of the standard connection 1-form $A$ \footnote{We use anti-hermitian generators 
of the gauge group, therefore no factors of $i$ appear 
in our expressions.}.
 
The form of the generalized field strength (\ref{gcurv}) 
is very suggestive. Based on it,
 it is possible to write down a generalized YM action 
on the Dirac structure
\be \label{GYM}
S_{\text{GYM}}=\int
\text{Tr}~\mc F\w\star_{L^{\star}}
\mc F~,
\ee
where the trace is taken over the non-abelian gauge group. 
The volume form
 and the Hodge duality operator refer 
to the Dirac structure. 
More precisely, the volume on the Dirac structure $L^{\star}$
is 
\bea 
\text{vol}_{L^{\star}}&=&
\text{vol}_{L^{\star}_{\parallel}}
\w\text{vol}_{L^{\star}_{\perp}}\nn\\
&=&\sqrt{-\text{det}g_{ab}}~\dd 
x^0\w\dots\w\dd x^p\w\nn\\ &&\w \sqrt{\text{det}g^{ij}}~ 
\partial_{p+1}\w\dots\w\partial_{m}\nn\\
&:=&\sqrt{-\text{det}g_{ab}} \sqrt{\text{det}g^{ij}}\dd
^{m+1} 
\mathcal{X}~,\label{vol}
\eea 
and it is part of the full generalized volume of 
${\mc T}\text{M}$.
In Eq. (\ref{vol}) we considered the splitting $L^{\star}=L^{\star}_{\parallel}\oplus L^{\star}_{\perp}$,
with $L^{\star}_{\parallel}\subset \text{T}^{\star}\text{M}$ 
and $L^{\star}_{\perp}\subset\text{TM}$. 
The Hodge operator takes a generalized $q$-vector 
of $\w^rL^{\star}_{\parallel}\w^{s}L^{\star}_{\perp}$ with 
$q=r+s$
and returns a dual generalized $(\text{dim}~\text{M}-q)$-vector of 
 $\w^{p+1-r}L^{\star}_{\parallel}\w^{m-p-s}L^{\star}_{\perp}$, 
such that $\star_{L^{\star}}\one=\text{vol}_{L^{\star}}$.
In order to make sense of the integration in Eq. (\ref{GYM}), 
we use a prescription similar to that of Ref. 
\cite{Ellwood:2006ya}  (see also Ref. \cite{Mylonas:2012pg}). 
In particular, we identify the measure on $L^{\star}_{\perp}$ 
with the measure $\dd^{m-p}p=\dd p_{p+1}\w\dots\w\dd p_{m}$ on a 
momentum space that is associated with the transverse coordinates 
$x^i$. This prescription makes sense only for fields that 
do not depend on $p_i$, which we assume from now on. 
This was already pointed out in Ref. \cite{Ellwood:2006ya} 
and it makes perfect sense e.g. for the case of D-branes, 
where the degrees of freedom depend only on the world volume 
coordinates. After this assumption is made, we are able to integrate out the $p_i$'s when necessary.
 This essentially corresponds 
 to dimensional reduction, albeit in an unusual guise. 
For the following we normalize this integrated measure 
to unity, namely $\int \sqrt{\text{det}g^{ij}}~ 
\dd p_{p+1}\w\dots\w\dd p_{m}=1$.

Using the expression (\ref{gcurv}), the action may be rewritten as 
 \bea\label{SGYM}
 S_{\text{GYM}}&=&\int
 \text{Tr}~\biggl(\sfrac 14 F\w\star F+\sfrac 12\sum_i D\Phi^i\w\star D\Phi^i+\nn\\&&\quad\qquad +\sfrac 14[\Phi^i,\Phi^j]^2
\text{vol}_{L^{\star}_\parallel}\biggl)\w\text{vol}_{L^{\star}_\perp}~,
\label{all}\eea
 where the gauge covariant derivative is
 \be\label{der}D_a=\partial_a+[A_a,~]~,\ee
and 
the Hodge duality operator $\star$ is now the standard one on 
$L^{\star}_{\parallel}$.
Then, if we start with $\dim \text{M}=10$, the action 
(\ref{GYM}) neatly reproduces 
some theories that are well-known. For 
$p=9$ it reduces to the bosonic sector of the maximally supersymmetric YM 
theory in ten dimensions,
while for $p=3$ one gets 
the bosonic sector of the ${\mc N}=4$ supersymmetric YM
 theory, after integrating out the $\text{vol}_{L^{\star}_{\perp}}$. 
We recall that in the 
 conventional formulation the ${\mc N}=4$ supersymmetric YM theory
in four dimensions is obtained by dimensional reduction of the $\mc {N}=1$ theory in ten 
 dimensions \cite{n4}. Here the action arises from generalized gauge theory on a specific Dirac structure and
 dimensional reduction is implemented by projecting 
the theory on $L^{\star}_{\parallel}$. 
Furthermore, for $p=-1$ the Dirac structure is 
the same as $\text{TM}$ and
we obtain an action 
that contains only the square of the scalar commutator, 
which is identical to the D-instanton matrix model 
\cite{ikkt}. 
Similarly, the D0 matrix model \cite{bfss} is obtained 
for $p=0$. 

The generalized field strength $\mc F$ is invariant
 under the generalized gauge transformation
\be\label{gaugetrafo}
\delta \mc A = \mc D\lambda = \dd_L\lambda + [\mc A,\lambda]~,
\ee
where $\lambda$ is a gauge transformation parameter. This reproduces exactly the expected gauge transformations for the gauge field $A$ and 
scalars $\Phi^i$ as we know them from the theories 
mentioned above. 

\section*{D-branes as Dirac structures.~}
The approach of the previous section exhibits a unified 
treatment of gauge and scalar fields. Therefore it is 
the appropriate framework to examine the D-brane effective 
action. 
Indeed, the authors of Ref. \cite{Satoshi} showed that a Dp-brane can be identified with a particular Dirac structure or, equivalently, 
with a specific leaf of a foliated structure.
In particular, the two dual Dirac structures $L$ and $L^{\star}$ that we introduced above were also introduced in Ref. \cite{Satoshi}
and they play a central role in their argument. 
The sections of the first one have the form
\be \label{spanl}
\mathfrak{X}_{L}=\upsilon^a(x)\partial_a+\xi_i(x)\dd x^i~,
\ee 
and $L$ contains all the necessary information for the geometric description of a D-brane. The second subbundle is the dual of $L$,
with sections 
\be 
\mathfrak{X}_{L^{\star}}=\upsilon^i(x)\partial_i+\xi_a(x)\dd x^a~,
\ee 
and it contains the information for the fluctuations
 of the brane, as well as for the world volume 
components of the background fields. The 
fluctuations of the brane world volume
 include the gauge field $A$ living on the brane, 
and the transverse scalars that form a vector $\Phi=\Phi^i(x)\partial_i$. This is suggestive of the generalized 
vector $\mc {A}=A+\Phi\in \G(L^{\star})$, which unifies the longitudinal and transversal degrees of freedom of the Dp-brane in a 
single mathematical object. As we saw above, it is part of the generalized connection (\ref{gc}) on the Lie algebroid $L$ with differential $\dd_L$ 
and curvature $\mc F$, given in Eq. (\ref{gcurv}).

Using the generalized field strength $\mc F$, one can 
determine the fluctuations of the brane just by considering the deformation of the Dirac structure $L$ corresponding to 
$$L_{\mc F}=e^{\mc F}L~.$$ The latter is also a Dirac structure 
provided that the intergrability condition 
\be 
\dd_L\mc F+\sfrac 12[\mc F,\mc F]_{S}=0
\ee 
holds \cite{wein}, where the bracket in this equation is the Schouten bracket.
The abelian case was considered in Ref. \cite{Satoshi}.
In the non-abelian case, which corresponds to multiple D-branes,
using Eqs. (\ref{gcurv}) and (\ref{spanl}) we find that 
\bea 
L_{\mc F}&=&\text{span}\{\partial_a+D_a\Phi^i\partial_i+F_{ab}\dd x^b,\nn\\
&&\quad\dd x^i-D_a\Phi^i\dd x^a+[\Phi^i,\Phi^j]\partial_j\}~,
\label{lf}\eea
with $D_a$ given in Eq. (\ref{der}). We observe that 
the commutator of the scalar fields appears in 
$L_{\mc F}$ and this will account for the 
non-abelian couplings of the D-brane.

\subsection*{The effective action for D-branes revisited}

In Ref. \cite{Satoshi} the generalized metric seen by the D-brane was 
determined and the DBI Lagrangian was reformulated in terms of it. According to our 
previous discussion on integration, the reformulated action takes the form
\be
-T_p\int
\dd^{10}\mathfrak X~ e^{-\phi} (\det g)^{1/4}~ (\det s_{\mc F})^{1/4}~,\label{dbi}
\ee
where $g$ is the Riemannian metric on $\text{TM}$ and
 $s_{\mc F} \in L^*\otimes L^*$ is the metric seen by the $\mc F$-twisted Dirac structure $L_{\mc F}$.
More precisely, denoting by $t$ the map from $L$ to
 $L^{\star}$ 
that implements T-duality geometrically,
 $s$ refers to its symmetric part 
and $a$ to its antisymmetric one, seen as an element of 
$\otimes^2L^{\star}$. Then the components of 
 $t$,
\be 
t_{ab}=E_{ab}-E_{ak}E^{kl}E_{lb}~,\quad t_{a}^{\ j}=
E_{ak}E^{kj}~,\quad t^{ij}=E^{ij}~,\nn
\ee 
 are identical to the ones of $E=g+B$ after 
a number of T-dualities, as they appear in Ref.
 \cite{Myers:1999ps}, where $B$ is the usual Kalb-Ramond 
2-form.
 Then $s_{\mc F}$ is determined 
to be equal to $s-(a-\mc F)s^{-1}(a-\mc F)$ \cite{Satoshi}. 
Although the action (\ref{dbi}) was formulated for a single D-brane, 
here we show that in the non-abelian case it remains 
the same, with the difference that the curvature 
$\mc F$ is now replaced by the non-abelian expression 
(\ref{gcurv}), and that it has to be traced over the gauge 
group.

In order to prove the above assertion, we essentially 
have to show that the product of the fourth roots of the 
determinants is equal to 
\be 
\sqrt{\frac{\text{det}\begin{pmatrix} E_{ab}-E_{ai}E^{ij}E_{jb}+F_{ab} & 
E_{ak}E^{kj}+D_a\Phi^j \\ -E^{ik}E_{kb}-D_b\Phi^i & E^{ij}+[\Phi^i,\Phi^j]
\end{pmatrix}}{\text{det}~E^{ij}}}~.\label{matrix}\nn
\ee 
 which is the expression for the
 full DBI Lagrangian \cite{Myers:1999ps}.
On the other hand the matrix that appears in this 
expression is equal to the components of the tensor $t_{\mc F}=
t-\mc F$. 
Furthermore, the authors of Ref. \cite{Satoshi} 
showed the following two identities in the abelian case,
\bea 
\text{det}~t_{\mc F}&=&(\text{det}~s)^{1/2}(\text{det}~
s_{\mc F})^{1/2}~,\label{id1}
\\
\text{det}~s&=&(\text{det}~E^{ij})^2\text{det}~g~.\label{id2}
\eea  
Following the same steps, a direct computation 
reveals that 
they also hold in the non-abelian case. The computation 
is unaltered for Eq. (\ref{id2}), which does not involve 
$A$ and $\Phi$ at all, and it may be found 
in Appendix A.6 of Ref. \cite{Satoshi}.
For Eq. (\ref{id1}) it is simple to see that 
the replacement of the abelian generalized curvature by 
the non-abelian counterpart (\ref{gcurv}) does not 
affect the calculation. 
Then it is simple to see that
\bea \mc L_{\text{DBI}}&=&
\sqrt{\sfrac{\text{det}~t_{\mc F}}{\text{det}~E^{ij}}}=
\sqrt{\sfrac{\sqrt{\text{det}~s}\sqrt{\text{det}~s_{\mc F}}}
{\text{det}~E^{ij}}}\nn\\&=&\sqrt{
\sqrt{\text{det}~g}
\sqrt{\text{det}~s_{\mc F}}}~,
\eea 
as required.

The remarkable property of the action (\ref{dbi}) is 
that it obviates the need to check that it 
is consistent with T-duality. 
Indeed, first it is important to observe that the factor 
\be 
\label{factor}
e^{-\tilde\phi}=e^{-\phi}(\text{det}~g)^{1/4}
\ee
is automatically invariant under T-duality. According to the 
usual Buscher rules for multiple T-dualities, 
\bea
e^{2\phi} &\to& e^{2\phi}\text{det}~E^{ij}~,\nn\\
g_{ab} &\to& g_{ab}~,\nn\\
g_{kl} &\to& \text{det}~g_{ij}~\text{det}~E^{ij}~g^{kl}~,
\eea
where we refer to the parametrization of Ref. \cite{Maharana:1992my}, 
in which the metric acquires the convenient 
factorization
$$
\text{det}~g=\text{det}~g_{ab}~\text{det}~g_{ij}~.
$$
It is then simple to see that $e^{-\tilde\phi}$ is T-duality invariant.
Now let us look at the remaining factor in the action, 
namely $(\text{det}~s_{\mc F})^{1/4}$. First of all, let 
us discuss the behavior of $\mc F$ under T-duality, since this 
will also be useful in the following. Splitting 
the world volume directions as $x^a=\{x^{\hat a},
x^{\hat i}\}$ and using the basic 
duality rule (\ref{basic}) for all directions labelled by $\hat i$, we obtain:
\bea 
\sfrac 12F_{ab}\dd x^{a}\w\dd x^b
&\to&\sfrac 12 F_{\hat a\hat b} 
\dd x^{\hat a}\w\dd x^{\hat b}+D_{\hat a}\Phi^{\hat i} \dd x^{\hat a}
\w \partial_{\hat i}\nn\\
&+&\sfrac 12[\Phi^{\hat i},\Phi^{\hat j}
]\partial_{\hat i}\w\partial_{\hat j}~,
\nn\\
D_{a}\Phi^i\dd x^a\w\partial_i&\to&D_{\hat a}\Phi^i\dd x^{\hat a}\w\partial_i+[\Phi^{\hat i},
\Phi^{i}]\partial_{\hat i}\w\partial_{i}~,\nn\\
\sfrac 12{[}\Phi^i,\Phi^j]\partial_{i}\w\partial_{j}&\to& \sfrac 12[\Phi^i,\Phi^j]\partial_{i}\w\partial_{j}~.\nn
\eea 
Summing up the respective sides of the above relations 
we directly obtain that the expression (\ref{gcurv}) for 
$\mc F$ does not change under T-duality. 
The same result holds for the character $e^{\mc F}$. 
On the contrary, recall 
that the Chern character $e^{F}$ does not behave well under 
T-duality \cite{Grange:2005nm}. 
Having proven the consistency of $\mc F$ with T-duality, it is 
easy to see that $s_{\mc F}$, and therefore the full 
action, is also consistent. 

A final essential remark has to do with the 
implementation of the gauge trace. In the conventional 
formulation one has to choose a prescription for this. 
In the present formalism, where all the expressions 
$F_{ab}, D_a\Phi^i$ and $[\Phi^i,\Phi^j]$ are components 
of the curvature of a single generalized connection, the symmetrized 
trace prescription, adopted in 
Ref. \cite{Myers:1999ps}, is automatic.

\subsection*{WZ Couplings}
We now turn our attention to 
the gauge invariant WZ couplings of D-branes to the RR fields.
 First we note that the generalized tangent bundle $\mc{T}\text{M}$ extends naturally to a Clifford bundle $Cl\text{M}$ using the 
non-degenerate bilinear operation in Eq. \eqref{bilinear} \cite{gg}. The Clifford bundle has a natural representation on 
 $\w^{\bullet} \text{T}^{\star}\text{M}$,
 given by the extension of 
the action
\be
\mathfrak{X}\cdot \omega = (\iota_X + \eta\w)\omega~,
\ee
where $\mathfrak{X} = X + \eta \in \G(\mc{T}\text{M})$ and
 $\omega\in \G(\w^{\bullet} \text{T}^{\star}\text{M})$. 
Then the generalized curvature $\mc F$ can be used to construct 
gauge invariant couplings to any section of
 $\w^{\bullet} \text{T}^{\star}\text{M}$. 
Let for example $C$ be a collection of odd p-forms,
given as the formal sum $C = \sum_{i=1}^9 C_i$, 
as in type IIA superstring theory. Simlar considerations hold 
in type IIB for even p-forms.
Consider the coupling 
\beq\label{WZnew}
\int \text{Tr}~\e^{\mc F}\cdot \left(C\w e^B\right)\w 
\text{vol}_{L^{\star}_{\perp}}~\big|_{L^\star}~,
\eeq
where the exponential is defined through its power 
series. Here we have introduced the restriction to 
$L^\star$ since otherwise the integrand will be a 
general section of $\w^\bullet{\mc T}\text M$. 
The spinorial actions of each term of 
$\mc{F}$ commute, which means that we can expand 
Eq. \eqref{WZnew} and obtain
\beq\label{WZold}
\int
\text{Tr}~\e^F\wedge \left[
\e^{D_a\Phi^i\dd x^a\wedge{\partial_i}}~\cdot~\e^{\iota_{\Phi}\iota_{\Phi}}~
C\w \e^B~\Big|_{L^\star}\right]~,
\eeq
where the action has been reduced to the world volume 
of the brane.
This is precisely the Wess-Zumino coupling of a stack 
of D-branes to the RR sector gauge potentials
 of type IIA string theory, 
as explained in Ref. \cite{Myers:1999ps} \footnote{It would be very interesting to include also the RR 
sector in the geometry (see e.g. Ref. \cite{jsv2}) and further simplify the WZ 
coupling.}.
 This becomes obvious using 
the fact that for any spacetime p-form $\omega$, 
\be\label{theorem1}
\text{P}\left[\omega\right] 
= \e^{D_a\Phi^i\dd x^a\wedge {\partial_i}}
\cdot\omega~\big|_{L^\star}~,
\ee
where on the left hand side we encounter the pullback of $\omega$ 
on the D-brane world volume. This assertion is easily 
proven by expanding the exponential and choosing a convenient gauge so that
\be 
\text{P}[\omega]_{ab}=\omega_{ab}+2\omega_{i[a}D_{b]}
\Phi^i+\omega_{ij}D_a\Phi^iD_b\Phi^j~,
\ee
where we have taken $\omega$ to be a 2-form for simplicity, 
but the result evidently holds for any p-form. Then using the relation (\ref{theorem1}), it is easily shown that 
the action (\ref{WZold}), and therefore the action (\ref{WZnew}) too, becomes
\be 
S_{\text{WZ}}=\int\text{Tr}~\e^{F}\w\text{P}\big[\e^{\iota_{\Phi}\iota_{\Phi}}~
C\w \e^B\big]~,
\ee
as required.

A final remark about the action (\ref{WZnew}) has to do with 
its behavior under T-duality. We already showed that 
$e^{\mc F}$ is consistent with T-duality. Moreover, 
it is well-known that although the RR gauge potentials 
$C_i$, as well as the sum $C$, are not well-behaved under 
T-duality, the improved ones $C'_i=C_i\w e^{B}$, and 
the corresponding sum, keep their form under it. Therefore 
we conclude that the action (\ref{WZnew}) is consistent 
with T-duality.

\subsection*{Discussion.~} 

In this letter, we studied non-abelian 
gauge theory in the framework of generalized complex 
geometry. We showed that this is the appropriate 
framework to discuss theories that involve standard 
gauge and scalar fields. Indeed, generalized gauge theory 
treats them on equal footing and unifies them in a single mathematical 
object, the generalized connection.
 The curvature of this 
connection can be used to define a generalized 
 Yang-Mills theory 
on a Dirac structure. It is remarkable that by changing 
Dirac structure this generalized YM theory scans a set 
of other standard theories, such 
as the ${\mc N=1}$ YM theory in 10D, the ${\mc N}=4$ 
YM theory in 4D, the IKKT and the BFSS matrix models.  
Moreover, the framework 
is also appropriate to discuss the world volume action 
of D-branes. Here we extended previous work that appeared in 
Ref. \cite{Satoshi}. In particular we argued that 
the abelian DBI action of Ref. \cite{Satoshi} is 
appropriate to describe the non-abelian case too, by 
promoting the generalized connection to its non-abelian counterpart.
Furthermore, we examined the WZ term and the non-abelian brane 
couplings and we showed that it acquires a very simple 
form in terms of the generalized curvature, while its 
consistency with T-duality is evident. 
 Although these expressions do not add anything 
substantially new from 
the physical point of view yet, the advantages of this 
rewritting are the following.
 First, the consistency of the action with T-duality 
can be checked with remarkable ease, essentially by 
simple inspection, since all the quantities that appear in it 
keep the same form under T-duality.
  Second, pullbacks and interior products do not appear 
 at all. The fields that enter 
the action are 
defined directly on spacetime and not on the world volume of 
the brane. 
In this sense, the action has more direct connection to the 
10D supergravity theory. Finally, the unification of 
gauge and scalar fields in a single generalized connection 
may turn out to be useful in physical contexts, 
such as in a unified description of dark 
energy and dark matter along the lines of Ref. \cite{Koivisto:2013fta} 
and in gaining new geometric insights for symmetry breaking 
in gauge theories.

\paragraph*{Acknowledgments.~}

 The authors are grateful to 
Larisa Jonke and Satoshi Watamura for discussions and for reading the manuscript, 
to Olaf Lechtenfeld for 
helpful remarks and feedback, and to Ivonne Zavala 
for discussions on a physical problem that inspired this 
work.

\end{document}